\documentstyle[12pt]{article}
\date{}
\author{Valerii Dryuma\thanks{Work supported in part by Grant RFFI, Russia-Moldova}\\[5mm]
{\it Institute of Mathematics and Informatics, AS RM,}\\[3mm] {\it
5 Academiei Street, 2028 Kishinev, Moldova},\\[3mm]{\it e-mail:
valery@dryuma.com; cainar@mail.md} }
\title{On solutions of Rashevskii equation}

\newtheorem{rem}{Remark}
\begin{document}
\maketitle
\date{}
\maketitle
\begin{abstract}
\ \ \ \ The solutions of Rashevskii equation for gonometric family
of plane curves are considered. Their properties are discussed.
The connection with the theory of duality for the second order
ODE's is discussed.
\end{abstract}

%\addc{ Author.}{ Title of abstract }
%\stopcr

\medskip
\section{Gonometric family of plane curves}

      Two parametrical family of plane curves is defined by the equation
      \begin{equation}\label{dr:eq03}
      F(x,y,\xi,\eta)=0.
      \end{equation}

      From the equation (\ref{dr:eq03}) and its differential  at
      fixed $(x,y)$
      \[
      F_\xi d \xi+F_\eta d \eta=0
      \]
can be find the coordinates
\begin{equation}\label{dr:eq06}
x=x(\xi,\eta,\frac{d\eta}{d\xi}),\quad
y=y(\xi,\eta,\frac{d\eta}{d\xi}).
\end{equation}

From the condition
\[
F_x dx+F_y dy=0
\]
is followed expression for the angle $\theta$
\[
\tan \theta=-\frac{F_x}{F_y}.
\]

After differentiating this expression one gets the equation for
the differential $d\theta$
\[
\frac{d \theta}{\cos \theta^2}=\frac{F_x\left(F_{y \xi}d\xi+F_{y
\eta}d \eta\right)-F_y\left(F_{x \xi}d\xi+F_{x \eta}d
\eta\right)}{F_y^2}.
\]
    Taking in consideration the formulaes (\ref{dr:eq06}) and the
    relation
\[
\cos ^2\theta =\frac{F_y^2}{F_x^2+F_y^2}
\]
 we get finally the expression
\begin{equation}\label{dr:eq1}
     {{\it d\theta}}^{2}=\frac{\left[\left(F_x F_{y\xi}-F_y F_{x\xi}\right)d\xi
     -\left(F_y F_{x\eta}-F_x F_{y\eta}\right)d\eta\right]^2}
     {\left(F_x^2+F_y^2\right)^2F_y^2}
\end{equation}
 which can be considered as the metric between two infinitely
    located in close proximity curves of a given family.

    In general case it has the form
    \[
    d\theta^2=\Phi(\xi,\eta,d\xi,d \eta).
    \]

    Two parametrical family of plane curves with angle
      metric having the Gauss form
      \[
      {{\it d \theta}}^{2}=g_{11}d \xi^2+2g_{12}d\xi d\eta+g_{22} d \eta^2
      \]
      is called gonometric.

     Two parametrical family of curves can be done in form of the second order ODE
     \begin{equation}\label{dr:eq02}
          y''=\left(1+y'^2\right)^{3/2}K(x,y,\phi),\quad \phi=\arctan y'
     \end{equation}
     where the function $K$ is the curvature along a curve from the family.

     The problem of description of two parametrical family of
     curves determined by the equations (\ref{dr:eq02})
     and having the Gauss angle form was solved by Rashevskii.

     As it was shown by Rashevskii the function $K(x,y,\phi)$ in the equation
     (\ref{dr:eq02}) must be solution of the partial differential equation
 \begin{equation}\label{dr:eq04}
\left(4K+2\sin \phi~ \partial_ x-2\cos \phi~ \partial_ y+\cos
\phi~
\partial_ x \partial_ \phi+\sin \phi
~\partial_ y \partial_ \phi+K\partial_ \phi \partial_
\phi\right)XK=0, \end{equation} where
\[
XK=\cos \phi \frac{\partial K}{\partial x}+\sin \phi
\frac{\partial K}{\partial y}+K\frac{\partial K}{\partial \phi}.
\]

     The aim of our consideration is integration of the equation (\ref{dr:eq04}).

\medskip
\section{The method of solution}

   For solutions of partial differential equation
 $$F(x,y,z,f_x,f_y,f_z,f_{xx},f_{xy},f_{yy},f_{xz},f_{yz},f_{zz},f_{xxx},=0$$
  we use the method of solution of
the p.d.e.'s  described first in \cite{dual} and developed later
in \cite{heav} .

      This method allow us to construct particular solutions of the partial nonlinear
       differential equation
\begin{equation}\label{Dr10}
F(x,y,z,f_x,f_y,f_z,f_{xx},f_{xy},f_{xz},f_{yy},f_{yz},f_{xxx},f_{xyy},f_{xxy},..)=0
 \end{equation}
    with the help of transformation of the function and corresponding variables.

     Essence of the method consists in a following presentation of the functions and
     variables
\begin{equation}\label{Dr11}
f(x,y,z)\rightarrow u(x,t,z),\quad y \rightarrow v(x,t,z),\quad
f_x\rightarrow u_x-\frac{u_t}{v_t}v_x,\]\[ f_z\rightarrow
u_z-\frac{u_t}{v_t}v_z,\quad f_y \rightarrow \frac{u_t}{v_t},
\quad f_{yy} \rightarrow \frac{(\frac{u_t}{v_t})_t}{v_t}, \quad
f_{xy} \rightarrow \frac{(u_x-\frac{u_t}{v_t}v_x)_t}{v_t},...
\end{equation}
where variable $t$ is considered as parameter.

  Remark that conditions of the type
   \[
   f_{xy}=f_{yx},\quad f_{xz}=f_{zx}...
   \]
are fulfilled at the such type of presentation.

  In result instead of equation (\ref{Dr10}) one get the
  relation between the new variables $u(x,t,z)$,~ $v(x,t,z)$ and
  their partial derivatives
\begin{equation}\label{Dr12}
\Psi(u,v,u_x,u_z,u_t,v_x,v_z,v_t...)=0.
  \end{equation}

    This relation coincides with initial  p.d.e at the condition $v(x,t,z)=t$
    and lead to the new p.d.e
  \begin{equation}\label{Dr13}
     \Phi(\omega,\omega_x,\omega_t,\omega_{xx},\omega_{xt},\omega_{tt},...)=0
    \end{equation}
     when the functions $u(x,t,s)=F(\omega(x,t,z),\omega_t...)$
    and $v(x,t,s)=\Phi(\omega(x,t,z),\omega_t...)$ are expressed through the auxiliary function $\omega(x,t,s)$.

     Remark that there are a various means to reduce the relation (\ref{Dr12}) into
    the partial differential equation.

    In some cases the solution of  equation (\ref{Dr13}) is a more simple
    problem than solution of the equation (\ref{Dr10}).
\begin{rem}

  As example we consider the  system of equations
    \[
   {\frac {\partial }{\partial z}}f(x,y,z)+f(x,y,z){\frac {\partial }{
\partial x}}f(x,y,z)+h(x,y,z){\frac {\partial }{\partial y}}f(x,y,z)
=0,
\]
and
\[{\frac {\partial }{\partial z}}h(x,y,z)+f(x,y,z){\frac {\partial }{
\partial x}}h(x,y,z)+h(x,y,z){\frac {\partial }{\partial
y}}h(x,y,z)=0.
\]

    Such type of equations are meeting in theory of
     motion of free fluid particles.

    After the $(u,v)$-transformation
    \[
    u(x,t)=\left ({\frac {\partial
}{\partial t}}\omega(x,t)\right )t- \omega(x,t),
\]
\[
v(x,t)=\left ({\frac {\partial }{\partial t}}\omega(x,t)\right )
\]
with condition
 \[
 \omega(x,t,z)=A(t,z)+x t
\]
it takes the form of  equation
\[
\left ({\frac {\partial }{\partial z}}A(t,z)\right
)^{2}+{t}^{2}\left ({\frac {\partial ^{2}}{\partial t\partial
z}}A(t,z)\right )^{2}-2\,t \left ({\frac {\partial ^{2}}{\partial
t\partial z}}A(t,z)\right ){ \frac {\partial }{\partial
z}}A(t,z)-\]\[-\left ({\frac {\partial ^{2}}{
\partial {t}^{2}}}A(t,z)\right ){t}^{2}{\frac {\partial ^{2}}{
\partial {z}^{2}}}A(t,z)
=0.
\]

The simplest  solution of this equation is
\[
A(t,z)={\it \_F1}(t)+{\it \_F2}(z)+{\it \_F2}(z){\it \_F1}(t) ,
\]
where \[ {\it \_F1}(t)=\left (\left ({\frac {{t}^{{{\it
\_c}_{{2}}}^{-1}}{\it \_c}_{{2}}}{\left ({\it \_C1}-{\it
\_C2}\,t\right )\left ({\it \_c}_{{ 2}}-1\right )}}\right
)^{{\frac {{\it \_c}_{{2}}}{{\it \_c}_{{2}}-1}}} \right )^{-1}-1,
\]
and
\[{\it \_F2}(z)=\left (\left (-{\it \_C1}\,z{\it \_c}_{{2}}+{\it \_C1}\,
z-{\it \_C2}\,{\it \_c}_{{2}}+{\it \_C2}\right )^{\left ({\it
\_c}_{{2 }}-1\right )^{-1}}\right )^{-1}-1
\]
with parameters $\it_\_C1,~ \it_\_C2,~\it_\_c_{2}$.

Elimination of the parameter ~$t$~ from
     corresponding relations give us the functions $f(x,y,z)$ and $h(x,y,z)$  satisfying the above
     system of equations.

     As example, in the case
     \[
     \it_\_C1=1,~ \it_\_C2=0,~\it_\_c_{2}=2
     \]
we get the solution
\[
h(x,y,z)=-1/2\,{\frac {-2\,{z}^{2}+\sqrt
{4\,yz-4\,xz}z-2\,yz+2\,xz}{{ z}^{2}}},
\]
and
\[
f(x,y,z)=-1/2\,{\frac {4\,yz-2\,\sqrt {4\,yz-4\,xz}z-4\,xz}{\sqrt
{4\, yz-4\,xz}z}}.
\]

    In the case
    \[
     \it_\_C1=1,~ \it_\_C2=0,~\it_\_c_{2}=1/2
     \]
we find
\[
h(x,y,z)=-{\frac {xz-yz+{x}^{2}-2\,xy+{y}^{2}-{z}^{2}}{{z}^{2}}}
\]
and
\[
f(x,y,z)=-{\frac {{x}^{2}-2\,xy+{y}^{2}-{z}^{2}}{{z}^{2}}}.
\]

    To construction of more complicated solutions it is necessary
    to use another type of reduction of $(u,v)$ -system.

\end{rem}
\medskip
\section{Abridged equation}

    The equation (\ref{dr:eq04}) admits  solutions satisfying the condition
\begin{equation}\label{dr:eq5}
XK=\cos \phi \frac{\partial K}{\partial x}+\sin \phi
\frac{\partial K}{\partial y}+K\frac{\partial K}{\partial \phi}=0.
\end{equation}

\subsection{Hodograph-transformation}

  To integrate the equation (\ref{dr:eq5}) we rewrite its in the
form
   \[
   \cos(w){\frac {\partial }{\partial u}}x(u,v,w)+\sin(w){\frac {
\partial }{\partial v}}x(u,v,w)+x(u,v,w){\frac {\partial }{\partial w}
}x(u,v,w)=0.
\]

Now transformations of the function and variables
\[
u-\lambda(x,v,w)=0,\quad {\frac {\partial }{\partial
u}}x(u,v,w)=\left ({\frac {\partial }{
\partial x}}\lambda(x,v,w)\right )^{-1},\]\[{\frac {\partial }{\partial v}}x(u,v,w)=-{\frac {{\frac {\partial }{
\partial v}}\lambda(x,v,w)}{{\frac {\partial }{\partial
x}}\lambda(x,v, ,w)}} ,\quad {\frac {\partial }{\partial
w}}x(u,v,w)=-{\frac {{\frac {\partial }{
\partial w}}\lambda(x,v,w)}{{\frac {\partial }{\partial x}}\lambda(x,v
,w)}}
\]
give us the linear equation on the function $\lambda(x,v,w)$
\[
\cos(w)-\sin(w){\frac {\partial }{\partial
v}}\lambda(x,v,w)-x{\frac {
\partial }{\partial w}}\lambda(x,v,w)=0
\]
with general solution
\[
\lambda(x,v,w)={\it \_F1}\left(x,-{\frac {\cos(w)+vx}{x}}\right
)+{\frac {\sin(w)} {x}},
\]
where $\it\_F1$ is arbitrary function of its own arguments.

  In result the function $x(u,v,w)$ is determined from the condition
  \[u-\lambda(x,v,w)=0,\]
  or
  \[
  u-{\frac {\sin(w)} {x}}={\it \_F1}\left(x,-{\frac {\cos(w)+vx}{x}}\right
).
\]

    As example  in the case
    \[
    {\it \_F1}(x,-{\frac {\cos(w)+vx}{x}})=x-{\frac {\cos(w)+vx}{x}}
\]
one get the equation
\[u-x+{\frac {\cos(w)+vx}{x}}-{\frac {\sin(w)}{x}}=0
\]
for determination the function $x=x(u,v,w)$.

   It is defined by the expression
\[
x(u,v,w)=1/2\,v+1/2\,u+1/2\,\sqrt
{{v}^{2}+2\,uv+{u}^{2}+4\,\cos(w)-4\,\sin(w)}.
\]

   From the correspondence
\[
 x(u,v,w)\Longleftrightarrow K(x,y,\phi)
\]
we find
\[
K(x,y,\phi)=1/2\,y+1/2\,x+1/2\,\sqrt
{{y}^{2}+2\,yx+{x}^{2}+4\,\cos( \phi)-4\,\sin(\phi)}
\]
the solution of abridged equation (\ref{dr:eq5}).

\subsection{(u,v)-transformation}

    From the sake of convenience we present the equation
    (\ref{dr:eq5}) in the form
    \begin{equation}\label{dr:eq6}
\cos(z){\frac {\partial }{\partial x}}K(x,y,z)+\sin(z){\frac {
\partial }{\partial y}}K(x,y,z)+K(x,y,z){\frac {\partial }{\partial z}
}K(x,y,z)=0. \end{equation}

  After application of the (u,v)-transformation at the equation (\ref{dr:eq6})
   we find the relation
   \begin{equation}\label{dr:eq7}
  \left (-\left ({\frac {\partial }{\partial t}}u(x,t,z)\right ){\frac {
\partial }{\partial z}}v(x,t,z)+\left ({\frac {\partial }{\partial z}}
u(x,t,z)\right ){\frac {\partial }{\partial t}}v(x,t,z)\right
)u(x,t,z )+\]\[+\left (\left ({\frac {\partial }{\partial
x}}u(x,t,z)\right ){\frac {\partial }{\partial t}}v(x,t,z)-\left
({\frac {\partial }{\partial t} }u(x,t,z)\right ){\frac {\partial
}{\partial x}}v(x,t,z)\right )\cos(z )+\sin(z){\frac {\partial
}{\partial t}}u(x,t,z)=0.
 \end{equation}

   By means of standard  change of variables
\[
v(x,t,z)=t\theta_{t}-\theta,\quad u(x,t,z)=\theta_{t}
\]
the relation (\ref{dr:eq7}) is reduced again to the nonlinear
equation
\[
\cos(z){\frac {\partial }{\partial x}}\theta(x,t,z)+\sin(z)+\left
({ \frac {\partial }{\partial t}}\theta(x,t,z)\right ){\frac
{\partial }{
\partial z}}\theta(x,t,z)=0.
\]

     In order to obtain an integrable reduction
     let us rewrite relation (\ref{dr:eq7})  in new designation as
     \begin{equation}\label{dr:eq8}
     \left (-\left ({\frac {\partial }{\partial v}}x(u,v,w)\right ){\frac {
\partial }{\partial w}}y(u,v,w)+\left ({\frac {\partial }{\partial w}}
x(u,v,w)\right ){\frac {\partial }{\partial v}}y(u,v,w)\right
)x(u,v,w )+\]\[+\left (\left ({\frac {\partial }{\partial
u}}x(u,v,w)\right ){\frac {\partial }{\partial v}}y(u,v,w)-\left
({\frac {\partial }{\partial v} }x(u,v,w)\right ){\frac {\partial
}{\partial u}}y(u,v,w)\right )\cos(w )+\]\[+\sin(w){\frac
{\partial }{\partial v}}x(u,v,w)=0. \end{equation}

   In result of application of hodograph-transformation
   \begin{equation}\label{dr:eq9}
y-\omega(x,v,w)=0,\quad u-\lambda(x,v,w)=0
\]
 the relation (\ref{dr:eq7}) takes the  form
\[\left (x{\frac {\partial }{\partial w}}\omega(x,v,w)-\sin(w)\right ){
\frac {\partial }{\partial v}}\lambda(x,v,w)+\left (-x{\frac {
\partial }{\partial w}}\lambda(x,v,w)+\cos(w)\right ){\frac {\partial
}{\partial v}}\omega(x,v,w)=0.
 \end{equation}

   From here it is easy to obtain the expressions
   \[
   \lambda(x,v,w)={\frac {\sin(w)}{x}}+{\it \_F1}(x,v)
\]
and
\[
\omega(x,v,w)=-{\frac {\cos(w)}{x}}+{\it \_F2}(x,v),
\]
where $ \_F2(x,v)$ and $ \_F1(x,v)$ are arbitrary.

    Now from the conditions (\ref{dr:eq9}) can be found the functions
    $x(u,v,w)$ and $y(u,v,w)$ and thereby the solutions of the equation
    (\ref{dr:eq6}).

    Let us consider an example.
\section{Simplest solutions of complete equation}

    In open form the Rashevskii equation has the form
    \begin{equation}\label{dr:eq10}
   1/2\,{\frac {\partial ^{3}}{\partial {y}^{2}\partial z}}K(x,y,z)+2\,
\sin(z)\left ({\frac {\partial }{\partial x}}K(x,y,z)\right
){\frac {
\partial }{\partial z}}K(x,y,z)-1/2\,\left ({\frac {\partial ^{2}}{
\partial {y}^{2}}}K(x,y,z)\right )\sin(2\,z)-\]\[-2\,\cos(z)\left ({\frac {
\partial }{\partial y}}K(x,y,z)\right ){\frac {\partial }{\partial z}}
K(x,y,z)+\left ({\frac {\partial ^{3}}{\partial y\partial
x\partial z} }K(x,y,z)\right )\sin(2\,z)+\]\[+2\,\cos(z)\left
({\frac {\partial ^{2}}{
\partial x\partial z}}K(x,y,z)\right ){\frac {\partial }{\partial z}}K
(x,y,z)+\cos(z)\left ({\frac {\partial }{\partial
x}}K(x,y,z)\right ){ \frac {\partial ^{2}}{\partial
{z}^{2}}}K(x,y,z)+\]\[+2\,\cos(z)K(x,y,z){ \frac {\partial
^{3}}{\partial z\partial x\partial z}}K(x,y,z)+2\,\sin (z)\left
({\frac {\partial ^{2}}{\partial y\partial z}}K(x,y,z)\right
){\frac {\partial }{\partial z}}K(x,y,z)+\]\[+\sin(z)\left ({\frac
{
\partial }{\partial y}}K(x,y,z)\right ){\frac {\partial ^{2}}{
\partial {z}^{2}}}K(x,y,z)+2\,\sin(z)K(x,y,z){\frac {\partial ^{3}}{
\partial z\partial y\partial z}}K(x,y,z)+\]\[+3\,K(x,y,z)\left ({\frac {
\partial }{\partial z}}K(x,y,z)\right ){\frac {\partial ^{2}}{
\partial {z}^{2}}}K(x,y,z)-1/2\,\left ({\frac {\partial ^{3}}{
\partial {y}^{2}\partial z}}K(x,y,z)\right )\cos(2\,z)+\]\[+\left (K(x,y,z)
\right )^{2}{\frac {\partial ^{3}}{\partial {z}^{3}}}K(x,y,z)+4\,
\left (K(x,y,z)\right )^{2}{\frac {\partial }{\partial
z}}K(x,y,z)- \]\[-\left ({\frac {\partial ^{2}}{\partial x\partial
y}}K(x,y,z)\right ) \cos(2\,z)+1/2\,\left ({\frac {\partial
^{3}}{\partial {x}^{2}
\partial z}}K(x,y,z)\right )\cos(2\,z)+1/2\,{\frac {\partial ^{3}}{
\partial {x}^{2}\partial z}}K(x,y,z)+\]\[+3\,K(x,y,z)\cos(z){\frac {
\partial }{\partial x}}K(x,y,z)+3\,K(x,y,z)\sin(z){\frac {\partial }{
\partial y}}K(x,y,z)+\]\[+1/2\,\left ({\frac {\partial ^{2}}{\partial {x}^{
2}}}K(x,y,z)\right )\sin(2\,z) =0.
\end{equation}

   Let us consider some simplest solutions (\ref{dr:eq10}).

   The substitution of the form
   \[
   K(x,y,z)={\frac {\sin(z)+1}{U(x)}},
\]
give us the equation for the function $U(x)$
\[
-5\,{\frac {d}{dx}}U(x)-\left ({\frac {d^{2}}{d{x}^{2}}}U(x)\right
)U( x)+3+2\,\left ({\frac {d}{dx}}U(x)\right )^{2}=0.
\]

    Its particular solutions are
    \[
    U(x)=x,\quad U(x)=\frac{3}{2}x.
    \]
   Remark that solution $U(x)=x$ corresponds the function
   \[
   K(x,y,z)={\frac {\sin(z)+1}{(x)}},
\]
which is also the solution of abridged equation, but the second
$U(x)=\frac{3}{2x}$ corresponds the solution of the complete
equation.

          The substitution of the form
\[
K(x,y,z)=U(x)\sin(z)+V(x)
\]
lead to the system of equations
\[3\,\left (U(x)\right )^{2}V(x)+4\,U(x){\frac {d}{dx}}V(x)+\left ({
\frac {d}{dx}}U(x)\right )V(x)+{\frac {d^{2}}{d{x}^{2}}}V(x)=0,
\]
\[
2\,{\frac {d^{2}}{d{x}^{2}}}U(x)+6\,U(x)\left (V(x)\right
)^{2}+4\,U(x ){\frac {d}{dx}}U(x)+6\,V(x){\frac {d}{dx}}V(x)=0
\]
having the particular solutions
    \[
    U(x)=\tanh({\frac {x+{\it \_C2}}{{\it \_C1}}}){{\it
    \_C1}}^{-1},\quad V(x)=0,
\]
and
\[
    U(x)=\tanh({\frac {x+{\it \_C2}}{{\it \_C1}}}){{\it
    \_C1}}^{-1},\quad  V(x)={\it \_C1}\,{e^{\int \!-U(x){dx}}}.
\]

       The substitution of the form
\[
K(x,y,z)=\cos(z){\frac {\partial }{\partial
y}}U(x,y)-\sin(z){\frac {
\partial }{\partial x}}U(x,y)
\]
give us the Liouville equation for the function $U(x,y)$
\[
{\frac {\partial ^{2}}{\partial {y}^{2}}}U(x,y)+{\frac {\partial
^{2}} {\partial {x}^{2}}}U(x,y)=M{e^{2\,U(x,y)}}.
\]

\section{Connection with dual equation}

   If the equation
   \[
   y''=\phi(x,y,y')
   \]
 is dual the second order ODE cubic on the first derivative
\[
b''=A(a,b)b'^3+B(a,b)b'^2+C(a,b)b'+E(a,b)
\]
then  its function $\phi(x,y,y')$  satisfies the p.d.e
    \begin{equation}\label{dr:eq11}
 \frac{d^2}{dx^2}\phi_{uu}-\phi{_u}\frac{d}{dx}\phi_{uu}-4\frac{d}{dx}\phi_{yu}+4\phi_{u}\phi_{yu}-3\phi_{y}\phi_{uu}+6\phi_{yy}=0,
 \end{equation}
here \[\frac{d}{dx}=\frac{\partial}{\partial
x}+u\frac{\partial}{\partial y}+\phi\frac{\partial}{\partial u}
\] and $u=y'$.

   It can be presented in form of the system
   \cite{dual},~ \cite{proj}
\[\phi_{xu}+\phi\phi_{uu}-\frac{1}{2}\phi_{u}^2+u\phi_{yu}-2\phi_{y}=h(x,y,u)
\]
\[
h_{xu}+\phi h_{uu}-\phi{_u}h_{u}+uh_{yu}-3h_{y}=0.
\]

    Such type of couple of equations
has common  General Integral $$
\begin{array}{ccccc}
 &  & F(x,y,a,b)=0 &  & \\
 & \swarrow \nearrow &  & \searrow \nwarrow & \\
y''=\phi(x,y,y') & & & &
b''=A(a,b)b'^3+B(a,b)b'^2+C(a,b)b'+E(a,b).
\\
\end{array}
$$

    In theory of the second order ODE having gonometric property its General
    Integrals
    \[
    F(x,y,a,b)=0
    \]
is also special - angle metric in the family of its curves has the
Gauss form
\[
d\theta^2=g_{11}(a,b)da^2+2g_{12}(a,b)dadb+g_{22}(a,b)db^2.
\]

Thereby the
   geodesic equations of such type metric are in the form
\[
b''=A(a,b)b'^3+B(a,b)b'^2+C(a,b)b'+E(a,b).
\]

     The relation between the equations
     \[
   y''=\phi(x,y,y')
   \]
   with condition (\ref{dr:eq11}) and
   the equations
   \[
    y''=(1+y'^2)^(3/2)K(x,y,z)
     \]
 with the function $K(x,y,z)$ satisfying the Rashevskii equation
(\ref{dr:eq10}) is the subject of our consideration.

      The equation (\ref{dr:eq11}) can be written as
        \begin{equation}\label{dr:eq12}
             A(x,y,z)\cos(2\,z)+B(x,y,z)\sin(2\,z)+C(x,y,z)\cos(z)+E(x,y,z)\sin(z)+F(x,y,z)=0,
\end{equation}
 where
\[
A(x,y,z)=\]\[=3\,{\frac {\partial ^{2}}{\partial
{y}^{2}}}K(x,y,z)-8\,{\frac {
\partial ^{3}}{\partial y\partial x\partial z}}K(x,y,z)-{\frac {
\partial ^{4}}{\partial z\partial {y}^{2}\partial z}}K(x,y,z)+{\frac {
\partial ^{4}}{\partial z\partial {x}^{2}\partial z}}K(x,y,z)-3\,{
\frac {\partial ^{2}}{\partial {x}^{2}}}K(x,y,z),
\]
\[
B(x,y,z)=-4\,{\frac {\partial ^{3}}{\partial {y}^{2}\partial
z}}K(x,y,z)+2\,{ \frac {\partial ^{4}}{\partial z\partial
y\partial x\partial z}}K(x,y, z)+4\,{\frac {\partial
^{3}}{\partial {x}^{2}\partial z}}K(x,y,z)-6\,{ \frac {\partial
^{2}}{\partial x\partial y}}K(x,y,z),
\]
\[
C(x,y,z)=4\,K(x,y,z){\frac {\partial ^{4}}{\partial
{z}^{2}\partial x\partial z }}K(x,y,z)+28\,K(x,y,z){\frac
{\partial ^{2}}{\partial x\partial z}}K( x,y,z)+\]\[+2\,\left
({\frac {\partial }{\partial x}}K(x,y,z)\right ){ \frac {\partial
^{3}}{\partial {z}^{3}}}K(x,y,z)+8\,\left ({\frac {
\partial }{\partial x}}K(x,y,z)\right ){\frac {\partial }{\partial z}}
K(x,y,z)-36\,K(x,y,z){\frac {\partial }{\partial
y}}K(x,y,z)-\]\[-6\,\left ({\frac {\partial }{\partial
y}}K(x,y,z)\right ){\frac {\partial ^{2}} {\partial
{z}^{2}}}K(x,y,z)+8\,\left ({\frac {\partial }{\partial z}}K
(x,y,z)\right ){\frac {\partial ^{2}}{\partial y\partial
z}}K(x,y,z)-\]\[-6 \,K(x,y,z){\frac {\partial ^{3}}{\partial
z\partial y\partial z}}K(x,y ,z)-2\,\left ({\frac {\partial
}{\partial z}}K(x,y,z)\right ){\frac {
\partial ^{3}}{\partial z\partial x\partial z}}K(x,y,z),
\]
\[E(x,y,z)=36\,\left ({\frac {\partial }{\partial x}}K(x,y,z)\right )K(x,y,z)-8\,
\left ({\frac {\partial }{\partial z}}K(x,y,z)\right ){\frac {
\partial ^{2}}{\partial x\partial z}}K(x,y,z)+\]\[+6\,\left ({\frac {
\partial }{\partial x}}K(x,y,z)\right ){\frac {\partial ^{2}}{
\partial {z}^{2}}}K(x,y,z)+4\,K(x,y,z){\frac {\partial ^{4}}{\partial
{z}^{2}\partial y\partial z}}K(x,y,z)+\]\[+6\,K(x,y,z){\frac
{\partial ^{3} }{\partial z\partial x\partial
z}}K(x,y,z)+28\,K(x,y,z){\frac {
\partial ^{2}}{\partial y\partial z}}K(x,y,z)-\]\[-2\,\left ({\frac {
\partial }{\partial z}}K(x,y,z)\right ){\frac {\partial ^{3}}{
\partial z\partial y\partial z}}K(x,y,z)+8\,\left ({\frac {\partial }{
\partial z}}K(x,y,z)\right ){\frac {\partial }{\partial y}}K(x,y,z)+\]\[+2
\,\left ({\frac {\partial }{\partial y}}K(x,y,z)\right ){\frac {
\partial ^{3}}{\partial {z}^{3}}}K(x,y,z),
\]
\[F(x,y,z)=\]\[=18\,\left (K(x,y,z)\right )^{3}+20\,\left (K(x,y,z)\right )^{2}{\frac
{\partial ^{2}}{\partial {z}^{2}}}K(x,y,z)+2\,\left
(K(x,y,z)\right )^ {2}{\frac {\partial ^{4}}{\partial
{z}^{4}}}K(x,y,z)+\]\[+9\,{\frac {
\partial ^{2}}{\partial {y}^{2}}}K(x,y,z)+{\frac {\partial ^{4}}{
\partial z\partial {y}^{2}\partial z}}K(x,y,z)+9\,{\frac {\partial ^{2
}}{\partial {x}^{2}}}K(x,y,z)+{\frac {\partial ^{4}}{\partial z
\partial {x}^{2}\partial z}}K(x,y,z).
\]

  So the problem is to find a common solutions of the equations
  (\ref{dr:eq12}) and (\ref{dr:eq10}).

    An existence  of such solutions show the example.

    The function
    \[
    K(x,y,z)={\frac {A\left (\sin(z)+1\right )}{x}}
    \]
 is the solutions of both  equations (\ref{dr:eq12}) and
 (\ref{dr:eq10}) at the conditions
 \[
 A=0,\quad A=1,\quad A=\frac{2}{3}.
 \]

    In the case $A=1$ a corresponding pair of ODE's looks as
    \cite{proj}
     \[
{\frac {d^{2}}{d{x}^{2}}}y(x)={\frac {\left ({\frac {d}{dx}}y(x)
\right )^{3}+{\frac {d}{dx}}y(x)+\left (\left ({\frac {d}{dx}}y(x)
\right )^{2}+1\right )^{3/2}}{x}}
\]
 and
\[
{\frac {d^{2}}{d{a}^{2}}}b(a)=-1/2\,{\frac {\left ({\frac
{d}{da}}b(a) \right )\left (\left ({\frac {d}{da}}b(a)\right
)^{2}+1\right )}{a}}.
\]

     Both equations have gonometric angle metric in space of their integral
     curves.

    The quantity  of such type of examples my be increased.

    In the case
    \[
    K(x,y,z)=A(x)\sin(z)+B(x)
\]
we get the system of equations for determination of $A(x)$ and
$B(x)$
\[
{\frac {d}{dx}}B(x)={\frac {B(x)\left ({\frac {d}{dx}}A(x)-\left
(A(x) \right )^{2}+\left (B(x)\right )^{2}\right )}{A(x)}}
\]
\[
{\frac {d^{2}}{d{x}^{2}}}A(x)=-{\frac {3\,\left (B(x)\right )^{2}{
\frac {d}{dx}}A(x)+3\,\left (B(x)\right )^{4}+2\,\left (A(x)\right
)^{ 2}{\frac {d}{dx}}A(x)}{A(x)}}.
\]
 In the case
    \[
  K(x,y,z)=\left ({\frac {\partial }{\partial y}}U(x,y)\right )\cos(z)-
\left ({\frac {\partial }{\partial x}}U(x,y)\right )\sin(z)
\]
which corresponds the equations cubic on the first derivative the
function $U(x,y)$ is solution of the Liouville equation.


\smallskip
\begin{thebibliography}{10}

\smallskip
\bibitem {sur}  V. Kagan, {\it Osnovy teorii poverhnostei}, v.2, OGIZ, Moskva,
(1948).

\smallskip
\bibitem {heav} V. Dryuma, {\it On solutions of the heavenly equations and their generalizations},
ArXiv:gr-qc/0611001 v1, 31 Oct 2006, pp.1-14.
\smallskip
\bibitem {dual} V. Dryuma, {\it On dual equation in theory of the second order ODE's},
ArXiv:nlin/0001047 v1 22 Jan 2007, pp.1-17.

\smallskip
\bibitem {cycl} V. Dryuma, {\it On geometry of gonometric family of cycles},
Arxiv:0710.181 nlin.SI v1 8 Oct 2007, pp.1-12.

\smallskip
\bibitem {proj} V. Dryuma, {\it Projective duality in theory of second order ODE's},
Mathetical Research, v.112, 1990, Kishinev, Stiinta, pp.93--103.

\smallskip
\bibitem {weyl} V. Dryuma, {\it The Riemann and Einsten-Weyl
geometries in theory of differential equations, their applications
and all that}. A.B.Shabat et all.(eds.), New Trends in
Integrability and Partial Solvability, Kluwer Academic Publishers,
Printed in the Netherlands , 2004, p.115--156.

\end{thebibliography}
\end{document}